\documentstyle[12pt,eqsecnum,preprint,aps]{revtex}
\title {Theory of double magnetophonon resonance \\ in 2D electron gas
in tilted magnetic field}
\author{{V. V. Afonin$^{a,b}$, V. L. Gurevich$^{a,b}$
R. Laiho$^{a}$}\\{\footnotesize\it $^{a}$Wihuri Physical Laboratory,
Department of Physics, University of Turku, FIN-20014 Turku,
Finland}\\{$^{b}$\footnotesize\it Solid State Physics Division, A.
F.  Ioffe Institute, 194021 Saint Petersburg,
Russia}}
\begin{document}
\maketitle
\begin{abstract}
A theory of double magnetophonon resonance (MPR) in quantum wells in
magnetic field is developed. The magnetic field is assumed to be
tilted at an angle $\theta$ to the perpendicular to the plane of
quantum well. The resonance is due to the resonant interaction of 2D
conduction electrons with the longitudinal optic phonons. The
electrons are assumed to be nondegenerate. The $\theta$-dependence of
MPR maxima is investigated. The existence of a double resonance, {\it
i.e.} two resonant peaks for each value of $\cal N$ (the number of
resonance) and the $\theta$-dependence of the MPR maxima is explained
by the screening in conjunction with the combined role of the phonon
and electron damping and variation of the electron concentration in
the well with magnetic field.

\end{abstract}

\section{Introduction}\label{IN}
Magnetophonon resonance (MPR) in semiconductors is an internal
resonance that is reached when the limiting frequency of a
longitudinal optic phonon equals the cyclotron frequency of an
electron, $\Omega$, times some small integer, $\cal N$ (for
instance,~\cite{R}).  Since its theoretical prediction~\cite{GF}
and subsequent experimental observation~\cite{PG,SPM} in
3D semiconductor structures MPR, along with cyclotron resonance,
has become one of the main instruments of semiconducting
compound spectroscopy.

The advances in semiconductor nano-fabrication in recent years have
made available materials of great crystalline perfection and purity.
The electrical conduction and some other transport phenomena in such
nanoscale structures has been a focus of numerous investigations,
both theoretical and experimental, with a number of important
discoveries. In particular, the discovery of MPR in the quantum wells
took place in the pioneering paper by Tsui, Englert, Cho and
Gossard~\cite{TECG}. After this first publication a number of papers
has appeared where various aspects of this physical phenomenon have
been investigated. The most detailed experimental investigation has
been done by Nicholas with co-workers (see the review paper~\cite{N}
and the references therein). It has been shown that the 2D MPR
qualitatively differs from the same phenomenon in the 3D structures.
As first steps in developing theory of 2D MPR in the perpendicular
magnetic field we can quote the theoretical papers~\cite{MMTH,DW}.
They consider the MPR within the perturbation theory approximation
with regard of collisional broadening of the electron state. A theory
in tilted magnetic field is developed in Ref.~\cite{SO} where
transitions to a higher band of spatial quantization are taken into
consideration. To our opinion, these theories do not provide detailed
interpretation of the existing experimental data.  Our purpose is to
give an interpretation of experimental findings~\cite{N}, such as the
double resonance, by simultaneously taking into account the screening
of the phonon potential by 2D conduction electrons as well as the
phonon and electron damping. For such a program, as we have indicated
in~\cite{AGL1}, the lowest approximation of the perturbation theory
is not sufficient. As is shown in the present paper, the relative
role of screening is determined not by the temperature (as has been
suggested in Ref.~\cite{BNH}) but by the interplay between the
screening and the electron and phonon collisional damping that is
usually weakly temperature dependent.

There are two main groups of the MPR experiments in quantum wells. The
first group deals with the MPR in the perpendicular (to the plane of
2DEG) magnetic field. The main features of the findings in this case
are ({\it i}) the fact that the resonance is determined by the
transverse optic frequency $\omega_t$ (rather than the longitudinal
frequency $\omega_l$) and ({\it ii}) a rather narrow interval of
electron concentrations where the MPR is observable. The second group
concerns with the experiments in a magnetic field tilted at an angle
$\theta$ to the perpendicular.  Its characteristic feature may be
called a {\em double resonance.} For small values of $\theta$ the MPR
is determined by $\omega_t$.  Then, for somewhat larger values of
$\theta$ its amplitude sharply goes down within a narrow angular
interval typically of the order of 10$^{\circ}$. For even bigger
values of $\theta$ there is another maximum, this time determined by
$\omega_l$~\cite{B1}.  These two types of resonance may be called the
$\omega_t$- and $\omega_l$-resonances as their positions are determined
by the frequencies $\omega_t$ and $\omega_l$ respectively.

In our paper~\cite{AGL1} we have given interpretation for the first
group of experiments. In the present paper we offer interpretation
of the second group. It is shown that the angular and concentration
dependencies of the MPR amplitudes are deeply interrelated.

The magnetic field $\bf{B}$ is assumed to be in the
$(y,z)$-plane, the $z$-axis being perpendicular to the 2DEG,
while the external electric field is oriented along the
$y$-axis. The $\rho_{yy}(=\rho_{xx})$ component of the
resistivity tensor will be calculated. This is the transport
coefficient expressed through the 2D conductivity
$\sigma_{\mu\nu}$ (averaged over the width of the well) as
$$\rho_{yy}=\sigma_{xx}/ (\sigma_{xy})^2.$$ As usual, it is
assumed that $\sigma_{xy}^2\gg\sigma_{xx}\sigma_{yy}$.  We consider
the situation where the well is so narrow that only one electron
band of spatial quantization is filled. Hence one can assume
that the $z$-component of electron velocity vanishes. Then
\begin{equation}
\dot{p}_x=eE_x+{eB\over c}v_y\cos\theta, \quad
\dot{p}_y=eE_y-{eB\over c}v_x\cos\theta
\label{1z}
\end{equation}
where $p_x$, $p_y$ and $v_x$  $v_y$ are respectively the components
of the electron quasimomentum and velocity. These classical equations
illustrate the physics describing a 2D circular motion of an electron
with the angular frequency
\begin{equation}
\Omega\cos\theta={eB\over mc}\cos\theta
\label{1t}
\end{equation}
where $m$ is the effective mass, so that $p_{x,y}=mv_{x,y}$. It follows
from Eq.~(\ref{1z}) that
\begin{equation}
\sigma_{xy}={enc\over B\cos\theta}
\label{2z}
\end{equation}
where $e$ and $n$ are the electron charge and concentration. It
means that within one miniband approximation the magnetic field
enters the problem only in the combination $B\cos\theta$.

\section{General equations}\label{GE}
To calculate the $x$-component of the d.c. current it is convenient
to consider the motion of a center of Landau oscillator. The
conductivity $\sigma_{xx}$ averaged over the width of the well
is given by (see Ref.~\cite{KHH})
\begin{equation}
\sigma_{xx}=\frac{e^2}{2k_{\rm B}TS}\int_{-\infty}^{\infty}dt
\left\langle\dot{X}(0)\dot{X}(t)\right\rangle
\label{1a}
\end{equation}
where $S$ is the area of the 2DEG, $T$ is the temperature, $X$ is the
operator of coordinate of the center of Landau oscillator in the
Heisenberg representation. According to Eq.~(\ref{1z}), in the
Schr\"odinger representation
$$ X=-{ic\over
eB\cos\theta}{\partial\over\partial y}+x
$$
({\it cf.} with~\cite{LL3}, Sec. 112).  It commutes with the free
electron Hamiltonian $\cal H$ in magnetic field $\bf B$ as well as
with the operator of Coulomb electron-electron interaction. This is a
consequence of the quasimomentum conservation in electron-electron
collisions. Here $\left\langle\dot{X}(0)\dot{X}(t)\right\rangle$ is
the ensemble averaged correlation function of velocities of
the centers of Landau oscillators. In the present and the following
sections we will usually assume $\hbar=1$, $k_{\rm B}=1$ and will
restore these symbols in the resulting formulas.

Now
\begin{equation}
\dot X(t)=\sum_{\sigma}\int\psi^{\dagger}({\bf r}, \sigma)
i[{\cal H}, X]\psi({\bf r}, \sigma)d^3r=
\sum_{\sigma}{c\over eB\cos\theta}\int\psi^{\dagger}({\bf r}, \sigma)
{\partial {\hat U}\over\partial y}\psi({\bf r}, \sigma)d^3r.
\label{2b}
\end{equation}
The summation is over the spin variable.
Here $\psi$ is the operator of the electron wave function while $\hat
U(t,{\bf r})$ is the operator of phonon field interacting with
electrons.  For the time being, we consider it as an external random
field; later on we will average over all its realizations introducing
the optic phonons.  The expression for $\sigma_{xx}$ can be presented
in such a form (we remind that we calculate the conductivity averaged
over the width of the well)
\begin{equation}
\sigma_{xx}=\frac{e^2}{2TS}\left(\frac{c}{eB\cos\theta}\right)^2
\int_{-\infty}^{\infty}dt
\int d^3r\int d^3r'
\left\langle {\hat n}(0,{\bf r}'){\partial\over\partial y'}
{\hat U}(0,{\bf r}'){\hat n}(t,{\bf r})
{\partial\over\partial y}{\hat U}(t,{\bf r})\right\rangle
\label{3x}
\end{equation}
where ${\hat n}(t,{\bf r})$ the electron density operator. Representing
the ensemble average in Eq.~(\ref{3x}) as a sum over the exact
quantum states of the system (see Ref.~\cite{LL9}, Sec. 36) we
get~\cite{er1}
\begin{eqnarray}
\sigma_{xx}&=&\frac{e^2}{2T}\left(\frac{c}{eB\cos\theta}\right)^2
\int\limits_{-\infty}^{\infty}{d\omega\over2\pi}\int\frac{d^2q}{(2\pi)^2}
\int dz\int dz'\frac{q_y^2N(\omega)}{1-\exp(-\omega/T)}\nonumber\\
&\times&[D_R(-\omega;{\bf q};z,z')-D_A(-\omega;{\bf q};z,z')]
[\Pi_R^{(3)}(\omega;{\bf q};z',z)-\Pi_A^{(3)}(\omega;{\bf q};z',z)]
\label{11x}
\end{eqnarray}
(see the details of the derivation in Ref.~\cite{AGL1}). We made use of
the quasimomentum conservation along the plane of the quantum well;
$\bf q$ is a 2D wave vector parallel to the plane of the well.
$N(\omega)$ is the Bose function. $\Pi_R^{(3)}(\omega;{\bf q};z',z)$ is
the exact 3D electron polarization operator.

Now, $D_R(\omega;{\bf q};z,z')$ is the phonon propagator with regard
of the direct Coulomb electron-electron (e-e) interaction [see below
Eq.~(\ref{3})]. In the present paper we consider the magnetophonon
resonant contribution to $\sigma_{xx}$. This means that the phonon
contribution to the Green function $D_R$ is determined by the optic
phonons. Further on we will assume that one can neglect the
difference between the lattice properties within and outside the
well. This assumption should not affect the qualitative results of
the theory. [Eq.~(\ref{11x}) permits to consider also a more general
(nonhomogeneous) case]. Without regard of e-e interaction the phonon
propagator has the form
\begin{equation}
D^{(0)}_{R,A}(\omega)=
{4\pi e^2\over (q^2+k^2)\varepsilon_c}
\cdot \frac{\omega_l}{2} \left(
\frac{1}{\omega - \omega_l \pm i \delta}-\frac{1}{\omega +
\omega_l \pm i \delta}\right)
\label{9x}
\end{equation}
where $k$ is the $z$-component of the wave vector while
$\varepsilon_c$ is given by
\begin{equation}
{1\over\varepsilon_c}={1\over\varepsilon_{\infty}}
-{1\over\varepsilon_0}.
\label{2}
\end{equation}
Here $\varepsilon_0$ and $\varepsilon_{\infty}$ are the dielectric
susceptibilities for $\omega\rightarrow0$ and
$\omega\rightarrow\infty$, respectively. We have included
the Fr\"olich electron-phonon interaction~\cite{F} into the definition
of the zero-order phonon Green function.

When calculating the exact phonon propagator it will be
necessary to insert along with the phonon lines $D^{(0)}_{R,A}$ the
direct Coulomb interaction lines
\begin{equation}
V^{(C)}({\bf q},k)={4\pi e^2\over\varepsilon_{\infty}(q^2+k^2)}.
\label{3}
\end{equation}
One should, however, observe the following important point. Both ends
of the exact phonon propagator {\it should be ordinary phonon lines
$D^{(0)}$ without Coulomb interaction lines.} This is due to the fact
that the operator $X$ commutes with the electron-electron interaction
operator [see Eq.~(\ref{2b})] as the latter conserves the electron
quasimomentum.

Further we assume for the electrons a parabolic confining potential
$m\omega_0^2z^2/2$ with the following gauge for the vector potential
${\bf{A}}=(-By\cos\theta + Bz\sin\theta, 0, 0)$. It is also assumed
that
\begin{equation}
\hbar\omega_0\gg\hbar\Omega, k_{\rm B}T
\label{1}
\end{equation}
(where $\Omega=eB/mc$ is the cyclotron frequency). In other words, we
assume the confining potential to be strong. The diagonalization
of quadratic Hamiltonian is a well-known procedure (for
instance, Ref.~\cite{M}). We wish, however, to emphasize that
the actual form of the confining potential is not essential for
the present theory provided that $\hbar\omega_l$ is much smaller
than the distance to the bottom of the second miniband [see
below --- Eq.~(\ref{1u})].

The energy of confined electron in the magnetic field defined by the
vector potential $\bf A$ is
\begin{equation}
{\cal U}={1\over2}m\Omega^2\cos^2\theta(y-y_0)^2-
{1\over2}m\Omega^2\sin(2\theta)z(y-y_0)+
{1\over2}m(\omega_0^2+\Omega^2\sin^2\theta)z^2
\label{12}
\end{equation}
where $y_0=-cp_x/eB\cos\theta$ while $p_x$ is the electron
quasimomentum component that is conserved. We will see in
Appendix
 that in the leading order in $(\Omega/\omega_0)^2$ one can
retain in Eq.~(\ref{12}) only the terms describing the electron
motion in the magnetic field $B\cos\theta$ perpendicular to the plane
of the well ({\em cf.} with Ref.~\cite{SH}). This can be visualized
in the following way. One can obviously neglect the magnetic field
correction to the confinement potential, i.e. $\Omega^2\sin^2\theta$
as compared to $\omega_0^2$.  This means that the characteristic
values of $z$ are of the order of $l=\sqrt{\hbar/\omega_0m}$. The
mixed term, i.e. the second term on the right-hand side of
Eq.~(\ref{12}) for the typical values of $z$ can be also discarded
provided that $\hbar\omega_0$ is the biggest energy in our problem.
Therefore the 2D polarization operator (calculated in detail in
Appendix) for the case of Boltzmann statistics and a small gas
parameter we are interested in has the same structure as the
polarization operator of Ref.~\cite{AGL1} in the perpendicular field
$B$ with the replacement $B\rightarrow B\cos\theta$:
\begin{equation}
\Pi_R=-2n_s\exp\left[-{(qa_B)^2\coth\alpha\over2\cos\theta}\right]
\sum_{{\cal N}={-\infty}}^{\infty} \frac{\sinh{\cal N}\alpha}
{\omega-{\cal N}\Omega\cos\theta+i\delta} \;I_{\cal N}
\left(\frac{q^2a_B^2}{2\cos\theta\sinh\alpha}\right).
\label{13}
\end{equation}
Here $I_{\cal N}$ is the modified Bessel function,
$\alpha=\hbar\Omega\cos\theta/2k_{\rm B}T$, $a_B^2=c\hbar/eB$,
$n_s$ is the 2D electron concentration. The polarization
operator $\Pi^{(3)}$ of Eq.~(\ref{11x}) differs from $\Pi_R$ by the
factor $\psi(z)^2\psi(z')^2$ due to the electron motion along
the $z$-axis. Here $\psi(z)$ is the wave function of the lowest
level of transverse quantization.

The e-e interaction can take place both via exchange of a phonon and
as a direct interaction described by $V^{(C)}$, Eq.~(\ref{3}). The
sum of two interactions is
\begin{equation}
V_{R,A}({\bf q},k)={4\pi e^2\over
(q^2+k^2)\varepsilon_{R,A}(\omega)},\quad
\varepsilon_{R,A}(\omega)= \varepsilon_{\infty}
\frac{\omega _{l}^{2}- (\omega\pm i\delta)^{2}} {\omega
_{t}^{2}-(\omega\pm i\delta)^{2}}.
\label{1y}
\end{equation}
Here $\omega_t^2=\omega_l^2(1-\varepsilon_{\infty }/\varepsilon_c)$.

Eq. (\ref{13}) shows that the electron-electron interaction cannot be
treated within the perturbation theory. Let the frequency $\omega$ in
Eq.~(\ref{13}) be close to the frequency ${\cal N}\Omega\cos\theta$
so that only one term of the series is important.  The higher orders
of the perturbation theory (without regard of the electron damping
$\Gamma_e$) give powers of an extra factor $1/ (\omega-{\cal
N}\Omega\cos\theta+i\delta)$.  Therefore, the e-e interaction must
be a sum of chains of loop diagrams (see Ref.~\cite{AGL1}).  Physically
this means taking into account the screening of phonon polarization
potential by the conduction electrons. Thus in 2D case in a resonance
the screening can be very important. The reason as to why one does not
need to take the screening into account in 3D case has been discussed
in Ref.~\cite{AGL1}. The only point demanding some attention is taking
into account the spatial nonhomogeneity. However, the procedure is
essentially facilitated by the fact that $\Pi(z_1,z_2)$ depends on
$\psi(z_1)^2\psi(z_2)^2$ as factors. Then the index of the progression
generated by the loops is proportional to
\begin{equation}
\int\!\int dz_1dz_2V_R({\bf q},z_1-z_2)\psi^2(z_1)\psi^2(z_2)
\label{1u}
\end{equation}
that can be presented in the form
$$
\int{dk\over2\pi}V_R({\bf q},k)\left[\int{ds\over2\pi}\psi_s
\psi_{s+k}\right]^2
$$ where $\psi_s$ is the Fourier component of $\psi(z)$.
This result is valid for a well of any form so far as the
distance to the bottom of the second miniband remains much
bigger than $\hbar\omega_l$. It is only necessary to insert into
Eq.~(\ref{1u}) the appropriate wave function $\psi(z)$.

For a quadratic confining potential we get a theory of 2D
electrons with the interaction potential
\begin{equation}
V_{R}(\omega,{\bf q})=\frac{4\pi
e^2}{\varepsilon_{R}(\omega)}\int_{-\infty}^{\infty}
{dk\over2\pi}{\exp(-k^2l^2/2)\over k^2+q^2}.
\label{14}
\end{equation}
Further on we will also need the expression
\begin{equation}
V^{(C)}({\bf q})=\frac{4\pi
e^2}{\varepsilon_{\infty}}\int_{-\infty}^{\infty}
{dk\over2\pi}{\exp(-k^2l^2/2)\over k^2+q^2}
\label{15}
\end{equation}
as well as the equation for the exact phonon Green function
\begin{equation}
{\cal D}_{R}(\omega,{\bf q})={V_{R}(\omega,{\bf q})\over1+
V_{R}(\omega,{\bf q})\Pi_{R}(\omega,{\bf q})}.
\label{16}
\end{equation}

Now we take into account the aforementioned point that both ends of
the chain in Eq.~(\ref{11x}) should be ordinary phonon lines (without
the Coulomb interaction). We have
\begin{equation}
D_R(\omega,{\bf q})=D^{(0)}_R(\omega,{\bf q})+
D^{(0)}_R(\omega,{\bf q}){1\over \Pi_R^{-1}-
V_{R}(\omega,{\bf q})}D^{(0)}_R(\omega,{\bf q})
\label{17}
\end{equation}
where
\begin{equation}
D^{(0)}_R(\omega,{\bf q})=\int {dk\over 2\pi}
D^{(0)}_R(\omega,{\bf q},k)\exp(-k^2l^2/2)
\label{19}
\end{equation}
and a purely 2D equation for the MPR
\begin{eqnarray}
\sigma_{xx}&=&\frac{1}{4T}\left(\frac{c}{B\cos\theta}\right)^2
\int\limits_{-\infty}^{\infty}{d\omega\over2\pi}\int\frac{d^2q}{(2\pi)^2}
\frac{q^2N(\omega)}{1-\exp(-\omega/T)}\nonumber\\
&\times&[D_R(-\omega;{\bf q})-D_A(-\omega;{\bf q})]
[\Pi_R(\omega;{\bf q})-\Pi_A(\omega;{\bf q})].
\label{18}
\end{eqnarray}
Here we made use of the fact that for $\hbar\omega_0\gg\hbar\Omega$
the integrand in Eq.~(\ref{18}) is symmetric in $q_x$ and $q_y$.

We will not insert directly Eq.~(\ref{17}) into Eq.~(\ref{18}) as it
seems to have poles at $\omega=\omega_l$ that in fact disappear after
integration and some algebra. It is convenient instead to present
Eq.~(\ref{17}) in the form
\begin{equation}
D_{R}(\omega,{\bf q})={\cal
D}_R(\omega,{\bf q})-V^{(C)}+\frac{2V^{(C)} V_R({\bf q})}{\Pi_R^{-1}+
V_R({\bf q})} -\frac{\left(V^{(C)}\right)^2} {\Pi_R^{-1}+ V_R({\bf q})}.
\label{20}
\end{equation}
One can see that neither of these terms has a pole $\omega=\omega_l$
[see Eq.~(\ref{14}) in combination with Eq.~(\ref{1y})]. This is a
manifestation of the influence of screening. It means that the
screening may play a certain role even for relatively small electron
concentrations.

The MPR is, as we will see, determined by the last term while the
contribution of all the rest terms in Eq.~(\ref{18}) vanishes (provided
that one neglects the electron and phonon damping). As a result, we
have for the $\cal N$th resonance of $\sigma_{xx}$
\begin{equation}
\sigma_{xx}=\frac{2n_s c^2\hbar^2} {\varepsilon_{\infty} k_{\rm
B}TB^2\cos^2\theta} {N(\omega_t)\sinh(\hbar\omega_t/2k_{\rm
B}T)\over1-\exp(-\hbar\omega_t/k_{\rm B}T)}{\cal J}_{\cal N}
\delta[{1/\varepsilon({\cal N}\Omega\cos\theta)}]
\label{21}
\end{equation}
where
\begin{equation}
{\cal J}_{\cal N}=\int_0^{\infty} dq q^3V^{(C)}(q) I_{\cal
N}\left(\frac{a_B^2 q^2}{2\cos\theta\sinh \alpha}\right)\exp
\left(-{a_B^2 q^2\coth\alpha\over2\cos\theta}\right)  .
\label{22}
\end{equation}
As is indicated in Sec.~\ref{IN}, it is natural that in the lowest
approximation in $\Omega^2/\omega_0^2$ only the combination
$B\cos\theta$ enters the equations describing the 2D motion of an
electron in the quantum well. The integrals in the rest terms in
Eq.~(\ref{20}) are either real or proportional to
$\varepsilon^{-1}\delta(\varepsilon^{-1})$. This can be easily
checked if one takes into consideration that
$\delta(\varepsilon^{-1})$ comes from the pole $[{\Pi_R^{-1}+
V_R({\bf q})}]^{-1}$ and the terms in question in Eq.~(\ref{20}) have
the factor $\varepsilon^{-1}$.

Further we will be interested in the case $ql\ll1$ when the
effective e-e interaction does not depend on the form of
potential and is equal to
\begin{equation}
V^{(C)}(q)={2\pi e^2\over\varepsilon_{\infty}q}.
\label{23}
\end{equation}
Then one can present ${\cal J}_{\cal N}$ as
\begin{equation}
{\cal J}_{\cal N}=
{4i\sqrt{\pi}e^2\over\varepsilon_{\infty} a_B^3}
\cos^{3/2}\theta\sinh^{1/2}
\left({\hbar\Omega\cos\theta\over2k_{\rm B}T}\right)
Q^1_{{\cal N}-1/2}\left(\cosh{\hbar\Omega\cos\theta
\over2k_{\rm B}T}\right)
\label{24}
\end{equation}
where $Q^1_{{\cal N}-1/2}(z)$ is the associated Legendre function of
the second kind (we remind that $Q^1_{{\cal N}-1/2}$ is an
imaginary function of a real argument).
We will consider the case $\alpha\ll1$. Then the characteristic
values of $q$ are of the order of $q_T=\sqrt{2mk_{\rm B}T}/\hbar$
and one can present ${\cal J}_{\cal N}$ in the following form
\begin{equation}
{\cal J}_{\cal N}={4\sqrt{2\pi}e^2\cos\theta\over
\varepsilon_{\infty}a_B^3}\sqrt{k_{\rm B}T\over\hbar\Omega}.
\label{25}
\end{equation}

As $\varepsilon^{-1}(\omega)$
[Eq.~(\ref{1y})] has a zero at $\omega=\omega_t$, $\sigma_{xx}$
exhibits in this approximation an infinitely narrow magnetophonon
resonance at
\begin{equation}
{\cal N}\Omega\cos\theta=\omega_t.
\label{26}
\end{equation}
Physically this is due to the fact that the
e-e interaction without regard of the damping is very strong in the
resonance.

\section{Angular dependence of MPR maxima}\label{AD}
In the present section we will investigate dependence of the
positions of the MPR maxima on the angle $\theta$. As we have
indicated, in the limit of vanishing phonon and electron damping
($\Gamma$ and $\Gamma_e$, respectively) the screening in the
resonance is very strong. If we take into account that the dampings
are finite one can calculate the critical concentration $n_s$ where
the screening ceases to play a role. As the interaction depends also
on $\theta$ for each value of $n_s$ one can indicate the
corresponding critical value(s) of $\theta$.

We will start with taking into account the phonon damping.
Finite optic phonon damping is due to the decay of an optic phonon
into two acoustic ones (see \cite{AGL1}). Technically it can be taken
into account by replacement $\omega\rightarrow\omega\pm i\Gamma$ in
the retarded and advanced phonon Green functions, respectively. One can
easily see that in such a case the MPR acquires a finite width which
one can take into account by the following replacement in
Eq.~(\ref{21})
\begin{equation}
\delta (\varepsilon^{-1})\rightarrow
{1\over\pi}{\rm Im}\, \varepsilon_R.
\label{14z}
\end{equation}
Here
\begin{equation}
{1\over\pi}{\rm
Im}\,\varepsilon_R={\omega_l^2-\omega_t^2\over2\pi\omega_t}
\frac{\varepsilon_{\infty}\Gamma}{\left({\cal N}
\Omega\cos\theta-\sqrt{\omega_t^2+\Gamma^2}\right)^2+\Gamma^2}.
\label{15z}
\end{equation}
In what follows we will assume that
\begin{equation}
\Gamma_e,\Gamma\ll\omega_l-\omega_t\ll\omega_t.
\label{27}
\end{equation}
One needs the first inequality to be able to discriminate between
frequencies $\omega_l$ and $\omega_t$. The second inequality is
fulfilled for such systems as GaAs/GaAlAs and facilitates
the calculations.

Now we will discuss the role of the electron damping. Good
examples of importance of the damping for the MPR are given in
Refs.~\cite{DW,D}. For us it is important as it may both destroy
the resonance and determine the angular interval for its
existence we are looking for.  We assume that
$\Gamma_e\ll\Omega\cos\theta$.  The electron Green function in
magnetic field has been investigated by Ando and
Uemura~\cite{AU} and in more detail by Laikhtman and E.
Altshuler~\cite{LA} for $\Gamma_e$ determined by an elastic
short range scattering. In the case we are particularly interested in,
{\em i.e.} GaAs the temperature variation of mobility from the liquid
helium temperature to the temperature of experiment (about 200K)
is substantial (in the typical cases, at least, by several
times). It means that the acoustic phonon scattering (that can
be considered as short range elastic) should be predominant.  It
was shown in Refs.~\cite{AU,LA} that in this case the electron
Green function has a non-Lorentzian form with the characteristic
width $\Gamma_e$ given by \begin{equation}
\Gamma_e=\Gamma_e^{(0)}\sqrt{\cos\theta}, \quad
\Gamma_e^{(0)}=\sqrt{\Omega/2\pi\tau}
\label{1f}
\end{equation}
where $\tau$ is the relaxation time for $B=0$ obtained by assuming the
same scatterers as for a finite $B$.

Further on our formulas should be considered as order-of-magnitude
estimates giving the correct dependencies on the parameters, though
not the parameter-independent numerical coefficients of the order of
unity.  For such estimates it will be sufficient to use the Lorentzian
form of $\Pi(\omega,{\bf q})$.  As has been indicated, in the resonance
approximation one should retain only the resonant term of all the
series~(\ref{13}) for $\Pi_R(\omega,{\bf q})$
\begin{equation}
\Pi_R(\omega,{\bf
q})=-\frac{{\cal R_{\cal N}(\omega,{\bf q})}} {\omega-{\cal
N}\Omega\cos\theta+i\Gamma_e}
\label{1e}
\end{equation}
where ${\cal
R}_{\cal N}$ is the residue at the pole $\omega={\cal
N}\Omega\cos\theta-i\Gamma_e$.  Calculating $\sigma_{xx}$ one can
evaluate the integral over the frequencies taking the residues in the
poles $\omega={\cal N}\Omega\cos\theta\pm i\Gamma_e$. To get the result
one should remove the factor $\delta(\varepsilon^{-1})$ in
Eq.~(\ref{21}) and insert instead into the integrand of Eq.~(\ref{22})
\begin{equation} \Delta\equiv{1\over\pi}{{\rm
Im}\,\varepsilon^{-1}_A+2\gamma\over(2\gamma+{\rm Im}\,
\varepsilon^{-1}_A)^2+({\rm Re}\,\varepsilon^{-1}_A)^2}
\label{2e}
\end{equation}
where
$$
\gamma={\Gamma_e\over\overline{\omega}};\quad \overline{\omega}={2\pi
e^2\over q}{\cal R}_{\cal N}({\cal N} \Omega\cos\theta,{\bf q})
$$
while $\varepsilon_A$ should be calculated at $\omega={\cal
N}\Omega\cos\theta+i(\Gamma+\Gamma_e)$. One can see that integral
(\ref{22}) is dominated by the values of $q$ where the asymptotic
expansion of the Bessel function is valid, so that
\begin{equation}
{\cal R}_{\cal N}={\sqrt{2}n_s\over\sqrt{\pi}}{{\cal N}\Omega\cos^2\theta
\over Tqa_B}\sqrt{\Omega\over T}\exp\left(-{q^2\over4q_T^2}\right).
\label{30}
\end{equation}

We remind the reader that Eq. (\ref{21}) is derived within the
so-called RPA (loop) approximation [see Eq. (\ref{17})] as it takes
into account the resonant interaction of electrons with optic
phonons. Such resonant terms should probably be added also to the
vertex parts describing the polaron effects. These terms may be
essential in the magnetophonon maxima. However, we are looking for
the critical values of the parameters such as electron concentration
where the resonant interaction disappears, so that the MPR signal
sharply goes down. In such a situation the polaron effects should be
also suppressed. Thus we believe that our theory, though probably not
permitting the exact calculation of the MPR amplitude, still gives
the correct characteristic values of the electron concentrations and
the angles where the MPR signal rapidly decreases.

In the present paper we limit ourselves with the sufficiently low
concentrations $n_s$ where the MPR maxima are well defined. As is
shown in Ref.~\cite{AGL1}, an additional mechanism of electron level
broadening due to the electron-electron interaction appears at high
electron concentration [see Eq. (6.10) of Ref.~\cite{AGL1})]. Here we
consider the concentrations that are not so large that one would have
to take into consideration this effect.

For small values of $\gamma$ Eq.~(\ref{2e}) turns into $\;{\rm
Im}\,\varepsilon_R(\omega_t)$ as one can neglect the terms $2\gamma$.
When $\gamma$ has reached the critical value
$(1/2)|\varepsilon^{-1}_A(\omega_t)|$ the amplitude of the maximum
begins to go down. This condition can be written as
\begin{equation}
2\gamma\approx {\rm Im}\,\left(\varepsilon_A\right)^{-1}.
\label{31}
\end{equation}
for
\begin{equation}
{\cal N}\Omega\cos\theta=\omega_t.
\label{31l}
\end{equation}
Under the MPR condition (\ref{31l}) ${\rm Im}\,
\left(\varepsilon_A\right)^{-1}$ is small. Therefore even
a small variation of the quantities determining $\gamma$
can violate Eq. (\ref{31}). The screening is important provided that
the parameter
\begin{equation}
\beta\equiv{\rm Im}\,\left[\varepsilon_A({\cal N}\Omega\cos\theta
+i\Gamma_e+i\Gamma)\right]^{-1}/2\gamma
\label{31a}
\end{equation}
is bigger than (or of the order of) unity. If it is much bigger than
unity the screening is strong and the result is independent of
$\beta$. The MPR maxima become $\beta$-dependent when $\beta$ is of
the order of unity. Then a relatively small variation of $\beta$
might drastically change the result (see Figs. 4 and 5 of
Ref.~\cite{AGL1}). To achieve such a change one needs a variation of
$\beta$ that need not be large.

Let us follow a variation of the MPR signal as a function of
$\theta$. One can write for an arbitrary magnetic field $B(\theta)$
that is near the $\cal N$th resonance value $B_{\cal N}(\theta)$
\begin{equation}
B(\theta)=B_{\cal N}(\theta)+\Delta B .
\label{31c}
\end{equation}
$B_{\cal N}(0)=\hbar\omega_tmc/e{\cal N}$ is the position of the 
${\cal N}$th MPR
maximum for $\theta=0$. The field corresponding to the $\cal N$th
resonance, varies for finite values of $\theta$ according to
\begin{equation} B_{\cal N}(\theta)\cos\theta=B_{\cal N}(0).
\label{31e}
\end{equation}
Thus the MPR amplitudes for various angles $\theta$ (and the same
resonance number $\cal N$) correspond to the same value of $\beta$
and therefore should coincide. (Only the width of the maximum should
enhance with decrease of $\theta$). This conclusion of the theory is
in drastic disagreement with the experiment~\cite{N}.

We think that this is due to the assumption that the electron
concentration $n_s$ is a constant independent of $B$. It is known,
however, that in the course of temperature variation from 77\,K to
300\,K at $B=0$ the variation of the carrier concentration may
comprise several tens per cent. This means that the energy variation
of some donors on the scale of the order of hundred K, or so
noticeably shifts the electron concentration balance between the
donors and the well.

Let us give a rough estimate of the variation of impurity level
positions as a function of $B$. If one assumes a hydrogen-like states
the variation $\delta\epsilon$ of their lowest level position
$\epsilon(B)$ under the shift of magnetic field $\delta B$ is (for
instance,~\cite{LL3}, section 112)
\begin{equation}
\delta\epsilon\approx-{\delta B\over B}
{\vert\epsilon(B)\vert\over2\ln[\hbar\Omega(B)/\epsilon_0]}
\label{31g}
\end{equation}
where $\epsilon_0={me^4/2\hbar^2\varepsilon_{\infty}^2}$ is the Bohr
energy while $\epsilon(B)$ is the position of the level in magnetic
field $B$, mark that $\epsilon(B)<0$.  This is a lower estimate as
with the overlap of atomic orbits the influence of the magnetic field
should enhance.  As in our case $\ln[\hbar\Omega(B)/\epsilon_0]$ is,
roughly, of the order of unity we will accept the estimate
$\delta\epsilon/\epsilon\approx-\delta B/B$. This should result in
redistribution of electrons between the donors and the well. In other
words, the electron concentration in the well will decrease (see
below). The chemical potential $\mu$ can be obtained from the
equation
\begin{equation} N_i=\exp\left({\mu\over k_{\rm B}T}\right)
\left\{N_i\exp\left[-{\epsilon_i(B)\over k_{\rm B}T}\right]
+{S\over2\pi a_B^2\cos^2\theta}\sum_L\exp
\left[-{\hbar\Omega_L(B\cos\theta)\over k_{\rm B}T}\right]\right\}
\label{31n}
\end{equation}
where $N_i$ is the total number of donors. Further on we will assume
that the number of electrons bound to the donors is much bigger than
in the well. This seems to be a rather typical experimental
situation. Practically the actual form of the impurity states and the
distribution of impurity levels may be (and usually is) much more
complicated. What is actually relevant to bring about a variation of
the electron concentration $n_s$ in the well is the magnetic field
dependence of the positions of impurity levels in the proper interval
of energies.

The variation of electron concentration for the values of $B$ given
by Eq. (\ref{31e}) is
\begin{equation}
{\delta n_s\over n_s(B)}=\xi{ \delta B\over B}\quad \mbox{with}\quad
\xi={\epsilon(B)/k_{\rm B}T\over1+n_s(B)/n_i(B)}
\label{31h}
\end{equation}
where $n_i(B)$ is the 2D concentration of the electrons bound by the
donors. It will be natural to assume below for the estimates that
$\vert\xi\vert$ is of the order of unity.

It is convenient to present $\beta$ as a ratio of the electron
concentrations to some characteristic value
\begin{equation}
\beta=n_s(B)/n_{\rm down}
\label{31i}
\end{equation}
where
\begin{equation}
{1\over n_{\rm down}}={\sqrt{2\pi}\hbar e^3\omega_t
B_{\cal N}(0)\over\
\varepsilon_{\infty}q_T T^2\Gamma_e mc}\cdot
{{\Gamma_e+\Gamma}\over{\omega_l-\omega_t}} .
\label{31j}
\end{equation}
$n_{\rm down}$ is the lower critical concentration where the
screening ceases to play a role. For $\cal N$=3 in GaAs its
characteristic value is about 10$^{10}$\,cm$^{-2}$. Equation~(\ref{31j})
differs from the equation for $n_{\rm down}$ given in
Ref.~\cite{AGL1} as we have assumed there $q_T\approx a_B^{-1}$
and $\hbar\omega_t\approx 2k_{\rm B}T$ that is valid for GaAs
under certain conditions and may differ for other situations.

The parameter $\beta$ depends on the electron concentration $n_s$.
This fact permits one to compare the dependence of the MPR amplitudes
in perpendicular magnetic field under the variation of concentration
(i.e., in different samples) and the angular dependence of the MPR
maxima under variation of $\theta$. This will permit to check as to
whether the magnetic field induced variation of $n_s$ is sufficient
to explain the decrease of the MPR amplitude as a function of
$\theta$ on the one hand and the existence of the subsidiary (side)
maximum in the angular dependence of the MPR amplitude on the other
hand (see Ref.~\cite{N}).

We start with discussion of the behavior of the MPR amplitude under
magnetic field rotation at small angles $\theta\ll1$. We will
consider the concentration interval $n_s\approx n_{\rm down}$ where
there is a sharp dependence of the MPR amplitude on $\beta$.
According to Eq.~(\ref{31e}) the MPR maximum will shift by
\begin{equation}
\delta B_{\cal N}\approx B_{\cal N}(0)(1-\cos\theta).
\label{31k}
\end{equation}
This will result in the relative variation of concentration $\delta
n_s/n_s\approx\xi(1-\cos\theta)$ that manifests itself in the angular
dependence of the MPR amplitude. As is known from the experiment in
the perpendicular magnetic field, in this region of concentrations
when $n_s$ goes down it brings about a sharp decease of the
amplitude.

As a reasoning supporting our view we will consider the following
numerical example.  According to Ref.~\cite{N}, Fig. 6 the variation
of the MPR amplitude is about 60\,\% provided the concentration varies
within the interval from 1.8$\cdot$10$^{10}$\,cm$^{-2}$ to
2$\cdot$10$^{10}$\,cm$^{-2}$. The same decrease of the amplitude
induced by magnetic field should take place for $\delta
n_s/n_s\approx 0.1$, i.e. for $\theta$ of the order of 20$^{\circ}$.
This shows an order-of-magnitude correspondence with the results of
Ref.~\cite{N}. Further increase of $\theta$ gets the system into the
region where $n<n_{\rm down}$ and a well defined resonant peak with
the resonant condition (\ref{31l}) disappears.

The behavior of this sort takes place provided that
$\beta{{\raise-8pt\hbox{$>$}}\atop\raise6pt\hbox{$\sim$}}1$.  For
such concentrations where $\beta\gg1$ small variation of this
parameter does not play an essential role. This means that for the
samples that have a maximal MPR amplitude
($n_s\approx5\cdot$10$^{10}$\,cm$^{-2}$) its $\theta$ dependence
should be absent.

Now we will treat the region of large angles.  Consider the samples
with the concentration of the order of $n_{\rm down}$. The angle
$\theta$ going up, $n_s$ can become so small that the perturbation
approach becomes applicable. In other words, one observes the
resonance determined by the condition
\begin{equation}
{\cal N}\Omega\cos\theta=\omega_l.
\label{31m}
\end{equation}
The screening is not important provided that $\beta<1$. In view of
Eq. (\ref{31m}) one should insert in (\ref{31a})
$\varepsilon_A(\omega_l+i\Gamma_e+i\Gamma)$. As a result, one can write
\begin{equation}
\beta_l={n_s(\theta)\over n_{\rm down}}\cdot\left({\omega_l-
\omega_t\over\Gamma_e+\Gamma}\right)^2.
\label{31o}
\end{equation}
If the last factor in this equation is large enough one needs rather
small concentrations $n_s(\theta)$. Let us estimate the angles where
they can be achieved. It follows from Eq. (\ref{31n}) under the same
assumptions provided that $\delta B/B$ is not small
\begin{equation}
n_s(\delta B+B)=n_s(B)\exp(\delta\epsilon/k_{\rm B}T).
\label{31p}
\end{equation}

Very rough estimates in the spirit of Ref.~\cite{FH} give [the
dependence $\epsilon(B)$ is rather smooth and we believe that
its expansion up to the linear term can be justified]
\begin{equation}
\delta\epsilon\approx-\zeta\epsilon_0\left({\hbar\Omega\over\epsilon_0}
\right)=-\zeta\hbar\Omega
\label{31q}
\end{equation}
where $\zeta$ is a number of the order of unity. This estimate is
based on the idea that the level shift depends on a single parameter,
i.e. the ratio of the magnetic energy to the Coulomb one (cf. the
analytical treatment of the hydrogen atom in magnetic field in
Ref.~\cite{LL3}, \S 112).

Thus the discussed $\omega_l$-maximum exists for the angles bigger
than $\theta_c$ given by the equation
\begin{equation}
\sec\theta_c\approx{1\over2}+\sqrt{{1\over4}+
{2{\cal N}k_{\rm B}T\over\zeta\hbar\omega_l}
\cdot\ln\left({\omega_l-\omega_t\over\Gamma_e+\Gamma}\right)} .
\label{31r}
\end{equation}
The second term under the sign of root is, most probably, of the
order of unity. In this case $\theta$ is somewhere in the interval
between 40$^{\circ}$ and 70$^{\circ}$, or so. Thus in the regions of
small and large angles we have well-defined $\omega_t$ and
$\omega_l$-resonances, respectively.  Their positions depend on
$B\cos\theta$. In the intermediate interval of angles there are no
well-defined MPR maxima.

It is interesting to note that in the region of small angles the width
of a maximum goes up with the number of resonance $\cal N$. It can be
easily seen if one takes into consideration enhancement of $n_s$
with decrease of magnetic field as well as enhancement of $n_c$ with
$\cal N$. All these conclusions are in a qualitative agreement with the
results described in Ref.~\cite{N}.

\section{Conclusion}

To summarize, we would like to stress that the interpretation of
behavior of the MPR in a tilted magnetic field has been a
long-standing problem~\cite{TECG,N,B1}. Two types of resonant
maxima have been discovered on experiment, {\em i.e.} the
$\omega_t$- and $\omega_l$-resonances as their positions are
determined by the frequencies $\omega_t$ and $\omega_l$,
respectively.

Important points to provide theoretical interpretation of these
resonances are the dependence of all quantities characterizing the 2D
motion of electrons on the combination $B\cos\theta$ whereas their
total concentration in the well depends on $B$. Had only the
$B\cos\theta$ dependence existed, the amplitude of the resonances,
i.e. the heights of the maxima would have been independent of the
angle $\theta$. We think that the only way to preserve the theory of
2D electron gas describing the concentration dependence of the MPR is
to assume variation of the electron concentration with $B$.  It
manifests itself in the angular dependence of the MPR amplitude. In
this way we have been able to give a qualitative interpretation of
the results given in Ref.~\cite{N}.

\section*{Acknowledgements}
V. V. A. and V. L. G. acknowledge support for this work by the
Academy of Finland, the Wihuri Foundation, and the Russian Foundation
for Basic Research, Grant No~00-15-96748.

\appendix
\section*{Calculation of polarization operator}\label{B}
As indicated in Sec.~\ref{GE}, we assume for the electrons a
parabolic confining potential $m\omega_0^2z^2/2$ with the gauge for
the vector potential ${\bf{A}}=(-By\cos\theta + Bz\sin\theta, 0, 0)$.
In spite of the first inequality~(\ref{1}), it is convenient to look for
the exact transformation of the Hamiltonian and solution of the
Schr\"odinger equation and only then go to the limit
\begin{equation}
\Omega/\omega_0\ll1.
\label{1b}
\end{equation}
Applying a standard procedure of diagonalization we get
for the bigger eigenvalue
\begin{equation}
\Omega_2^2=\omega_0^2+\Omega^2-\Omega_c^2
\label{2c}
\end{equation}
while the smaller eigenvalue is
\begin{equation}
\Omega_c^2={\Omega^2+\omega_0^2\over2}\left[1-\sqrt{1-{4\Omega^2
\omega_0^2\cos^2\theta\over(\Omega^2+\omega_0^2)^2}}\right].
\label{3b}
\end{equation}

The variables ($Y,Z$) diagonalizing the Hamiltonian are expressible
through the initial variables ($y,z$) as
\begin{equation}
\left({Y\atop Z}\right)={1\over C}\left(y-y_0-rz\atop r(y-y_0)+z\right).
\label{4b}
\end{equation}
Here $C={1/\sqrt{1+r^2}}$,
$$
r={\Omega_c^2-\Omega^2\cos^2\theta\over\Omega^2\cos\theta\sin\theta}.
$$
As a result, we get two noninteracting oscillators, {\em i.e.} the
$Y$-oscillator and the $Z$-oscillator
\begin{equation}
{\cal H}=-{1\over2m}{\partial^2\over\partial Y^2}-
{1\over2m}{\partial^2\over\partial Z^2}+{m\over2}\Omega_c^2Y^2+
{m\over2}\Omega_2^2Z^2.
\label{5b}
\end{equation}
We are interested in the eigenfrequency of the $Y$-oscillator that is
\begin{equation}
\Omega_c=\Omega\cos\theta
\label{6b}
\end{equation}
plus small terms proportional to $\Omega^2/\omega_0^2\ll1$. The
eigenfrequency of the $Z$-oscillator is equal to $\omega_0$ (with the
same accuracy).

Now we turn to calculation of the polarization operator for
nondegenerate free electrons in magnetic field by a slight
modification of the method proposed by Sondheimer and
Wilson~\cite{SW}. The method is based on the spectral representation
(see Sec. 36~\cite{LL9}).
\begin{equation}
\Pi_R(\omega,z',z)=\sum_{m,l}w_l\frac
{\left\langle m\vert\hat{n}(0,z')\vert l\right\rangle
\left\langle l\vert\hat{n}(0,z)\vert m\right\rangle}
{\omega-\omega_{lm}+i\delta}
\cdot [1-\exp(-\omega_{lm}/T)].
\label{7x}
\end{equation}

Sondheimer and Wilson introduced a Green function of a complex time
argument $\gamma$
\begin{equation}
G({\bf r},{\bf r}',\gamma)=\sum_{\beta}\psi^*_{\beta}({\bf r}')
\psi_{\beta}({\bf r})\exp\left(-\varepsilon_{\beta}\gamma\right),
\quad{\rm Re}\,\gamma>0.
\label{7b}
\end{equation}
Here $\beta$ is the set of all quantum numbers of an electron, while
$\varepsilon_{\beta}$ is the energy of the quantum state $\beta$. One
can express the polarization operator through the Green functions
of a complex time argument.

The retarded polarization operator for Hamiltonian~(\ref{5b}) is
\begin{eqnarray}
\Pi^{(3)}_R({\bf r},{\bf r}',t)&=&2i\cosh{\Omega_2+\Omega_c\over2T}
\exp\left({\mu\over T}\right)\Theta(t)\nonumber\\
&\times&\left[G\left({\bf r}',{\bf r},{1\over
T}-it\right) G({\bf r},{\bf r}',it)-G({\bf r}',{\bf r},-it) G\left({\bf
r},{\bf r}',{1\over T}+it\right)\right].
\label{8b}
\end{eqnarray}
Here $\psi_{\beta}({\bf r})$ is a product of the eigenfunctions of
the $Y$- and $Z$-oscillators, $\Theta(t)$ is the step function.

In order to sum up the series in Eq.~(\ref{7b}) over
the $Y$- and $Z$-oscillator quantum numbers we will use the following
relation~\cite{B}
\begin{eqnarray}
\sum\limits_0^{\infty}{1\over2^nn!}\exp\left[-{1\over2}(y^2+{y'}^2)
-n\gamma\omega\right]H_n(y)H_n(y')\nonumber\\
=\left(1-e^{-2\gamma\omega}
\right)^{-1/2}\exp\left[-{1\over4}(y+y')^2{{1-e^{-\gamma\omega}}\over
{1+e^{-\gamma\omega}}}-{1\over4}(y-y')^2{{1+e^{-\gamma\omega}}\over
{1-e^{-\gamma\omega}}}\right]
\label{9b}
\end{eqnarray}
where $H_n(y)$ is the Hermite polynomial. It is convenient to replace the
summation over the quantum number $p_x$ by integration over $y_0$.

Now,
$$
G({\bf r},{\bf r}',\gamma)={ a_1C eB\cos\theta\over \sqrt{\pi}a_2}
\biggl
(\sinh\gamma\Omega_c \sinh\gamma\Omega_0
\biggr )^{-1/2}
\biggl
(\tanh{\gamma\Omega_c\over2} +
{a_1^2r^2\over a_2^2}\tanh{\gamma\Omega_0\over2}
\biggr)^{-1/2}
$$
$$
\times{\rm exp}
\Biggl\{-{1\over4}\coth\biggl({\gamma\Omega_c\over2}\biggr)(\Delta Y)^2
-{a^2_1\over4a^2_2}\coth\biggl({\gamma\Omega_0\over2}\biggr) (\Delta
Z)^2-{1\over4}b_1^2 \tanh\biggl({\Omega_0\gamma\over2}
\biggr)(z+z')^2 $$
\begin{equation}
-{1\over4}
{C^2\bigl[\Delta
x+ib\tanh{(\Omega_0\gamma/2)}(z+z')\bigr]^2\over{\displaystyle
\tanh{(\gamma\Omega_c/2)} + {(a_1^2r^2/
a_2^2)}\tanh{(\gamma\Omega_0/2)}}} \Biggr\}.
\label{10b}
\end{equation}
Here
\begin{equation}
b={a_1^2r\over a_2^2},\quad b_1={b\over r};
\label{11b}
\end{equation}
the variables in Eq.~(\ref{10b}) are made dimensionless by the
transformations
\begin{equation}
\Delta Y=(Y-Y')/a_1,\quad\Delta Z=(Z-Z')/a_1,
\quad z\rightarrow z/a_1, \quad z'\rightarrow z'/a_1
\label{12b}
\end{equation}
where
\begin{equation}
a_1^2=1/m\Omega_c,\quad a_2^2=1/m\Omega_0.
\label{13b}
\end{equation}

To calculate the polarization operator it is convenient to use the
Green functions in the momentum representation. After rather involved
but straightforward calculations we get
\begin{equation}
\Pi^{(3)}({\bf q},q_z,Q_z,t)=P({\bf q},q_z,Q_z,t)+
P^*(-{\bf q},-q_z,Q_z,t)
\label{15b}
\end{equation}
where the asterisk means a complex conjugation;
\begin{eqnarray}
P({\bf
q},q_z,Q_z,t)=\frac{2iCa_1B^2\cosh{\displaystyle{\Omega_c
+\Omega_0\over2T}}\Theta(t)e^{\mu/T}}{\pi a_2\sinh({\Omega_c/2T})
\sinh({\Omega_0/2T})}
\exp\left\{-\left(q_y^2+{q_x^2\over
C^2}\right)g_1(\Omega_c) \right.\nonumber\\
-\left.\left.\left(q_z^2+{r^2q_x^2\over C^2}\right)
{a_2^2\over a^2_1}g_1(\Omega_0)
-{Q_z^2\over4C^2}\left[{a_2^2\over a_1^2}g_2(\Omega_0)+{r^2\over2}
g_2(\Omega_c)\right]
+{rq_xQ_z\over2C^2}[g_3(\Omega_c)-g_3(\Omega_0)]\right]\right\}
\label{14b}
\end{eqnarray}
where
$$
g_1(\Omega)={1\over2}\left[\coth{\Omega\over2T}-\cos\left(\Omega t+
{i\Omega\over2T}\right){\left/\sinh{\Omega\over2T}\right.}\right],
$$
$$
g_2(\Omega)={1\over2}\left[\coth{\Omega\over2T}+\cos\left(\Omega t+
{i\Omega\over2T}\right){\left/\sinh{\Omega\over2T}\right.}\right],
$$
$$
g_3(\Omega)={\sin\Omega(t+i/2T)\over\sinh(\Omega/2T)}.
$$
Here $\bf q$ and $q_z$ refer to the Fourier components over the
differences ${\bf r}-{\bf r}'$ and $z-z'$ while $Q_z$ is related to
$(z+z')/2$.

We are looking for a frequency representation of the polarization
operator. We are going to take into consideration only the lowest level
of transverse quantization, {\em i.e.} the lowest miniband. At the
first sight one could try to average Eq.~(\ref{14b}) over the time
interval bigger than $1/\omega_0$ but smaller than $1/\Omega\cos\theta$.
However, some spurious terms can appear as a result of the direct
averaging due to the even powers of $\sin\omega_0t$ and $\cos\omega_0t$.
Therefore we will look for the frequency representation of the whole
expression Eq.~(\ref{14b}). One can find it with the help of the
identities
\begin{equation}
\exp(-z\cos A)=\sum^{\infty}_{-\infty}e^{inA}I_n(z),\quad
\exp(-z\sin A)=\sum^{\infty}_{-\infty}e^{inA}J_n(iz).
\label{17b}
\end{equation}
where $J_m$ are the ordinary Bessel functions.
Combining them with Eq.~(\ref{14b}) one can see that $A$ is a linear
function of time, so that one can easily calculate the Fourier
components. Discarding the poles describing the transitions to the
higher minibands one gets
\begin{eqnarray}
\Pi_R^{(3)}&=&-2Cn_s\exp\left[-
\left({q_y^2\over2}+{q_x^2\over2C^2}\right)
\coth{\Omega_c\over2T}-{Q_z^2\over C^2}
\left({a_2^2\over8a_1^2}+{r^2\over8}\coth{\Omega_c\over2T}\right)-
{a_2^2\over a_1^2}
\left({q_z^2\over2}+{r^2q_x^2\over2C^2}\right) \right]\nonumber\\
&\times&\sum\limits_{lmn}I_{n+l}\left({q_y^2+q_x^2/C^2\over2
\sinh(\Omega_c/2T)}\right)
I_{l-m}\left({r^2Q_z^2\over8C^2
\sinh(\Omega_c/2T)}\right)\nonumber\\
&\times&J_m\left({irq_xQ_z\over2C^2\sinh(\Omega_c/2T)}\right)
{\sinh(\Omega_cn/2T)\over\omega-\Omega_cn+i\delta}.
\label{18b}
\end{eqnarray}

There are elegant formulas to calculate this sum. However, as we are
interested in the limiting case $\Omega^2/\omega_0^2\ll1$ one can
see the result immediately from Eq.~(\ref{18b}). Indeed, the Bessel
functions having in their arguments as a factor the small parameter $r$
can be discarded unless $l-m=m=0$ and we are left with
\begin{eqnarray}
\Pi_R^{(3)}&=&-2n_s\exp\left[-
{q_x^2+q_y^2\over2}\coth{\Omega_c\over2T}
-{a_2^2\over a_1^2}{Q_z^2\over8}-{a_2^2\over a_1^2}
{q_z^2\over2}\right]\nonumber\\
&\times&\sum\limits_{n}I_{n}
\left({q_y^2+q_x^2\over2
\sinh(\Omega_c/2T)}\right){\sinh\left(n\Omega_c/2T\right)
\over\omega-\Omega_cn+i\delta}.
\label{19b}
\end{eqnarray}
One can see that $\Pi^{(3)}$ Eq.~(\ref{19b}) coincides with
$\Pi_R^{(3)}$ of Eq.~(\ref{11x}) and (\ref{13}) if one takes into
consideration that within the accepted
approximation $\Omega_c=\Omega\cos\theta$ and that we use in
Eq.~(\ref{19b}) dimensionless variables.


\begin{thebibliography}{99}
\bibitem{er1}In Eq. (3.13) of Ref.~\cite{AGL1} we erroneously
omitted the factor $N(\omega_l)/[1-\exp(-\omega_l/T)]$ while in
Eqs. (5.8) and (5.9) the factor $N(\omega_t)/[1-\exp(-\omega_t/T)]$
is omitted.
\end{thebibliography}
\end{document}